\documentclass{article}
\usepackage{graphicx}
\usepackage{amsmath}
\usepackage{amsfonts}
\usepackage{amssymb}
\usepackage{braket}
\usepackage{mathrsfs}
\usepackage{comment}
\usepackage{tikz}
\usepackage{jheppub}

\title{Remarks on Associated Varieties and Minimal Tension Holography}

\subheader{\normalsize
        \vspace*{-3.7em}
        \begin{flushright}
	DESY-25-072
        \end{flushright}
} 

\date{December 2024}

\newcommand{\omin}{ \bar{\mathbb{O}}_{\mathrm{min}}(\mathfrak{sl}_n^*)}

\begin{document}

\author[1,2,3]{Andrea E. V. Ferrari}

\affiliation[1]{Deutsches Elektronen-Synchrotron DESY, Notkestr. 85, 22607 Hamburg, Germany}
\affiliation[2]{School of Mathematics, The University of Edinburgh, Mayfield Road,  EH9 3JZ Edinburgh, U.K.}
\affiliation[3]{DAMTP, University of Cambridge, Wilberforce Rd, CB3 0WA Cambridge, UK}

\emailAdd{andrea.e.v.ferrari@gmail.com}

\abstract{We comment on certain analogies that have recently emerged between path integral localisation phenomena in minimal tension string theory on $\mathrm{AdS}_3\times S^3\times T^4$ and associated varieties of boundary vertex algebras of 3d $\mathcal{N}=4$ theories. We give evidence for the fact that the path integral localisation relies on an intriguing relation between the conformal boundary of $\mathrm{AdS}_3$ and the associated variety of $V_1(\mathfrak{psl}(2|2))$, (the complexification of) the affine algebra that enters the world-sheet description of the string theory. We explain the origin of certain operators (related to ``secret representations") that have appeared in this description. Based on the expectation that similar statements will hold for $\mathrm{AdS}_5\times S^5$, we explicitly write down free field realisations of $V_1(\mathfrak{psl}(4|4))$ that may hopefully be useful to study the world-sheet dual of free $\mathcal{N}=4$ SYM.}

\maketitle

\section{Introduction and motivation}

The holographic correspondence between string theory on $\mathrm{AdS}_3 \times S^3 \times T^4$ in the minimal tension ($k=1$) limit and the large-$N$ free symmetric product orbifold of CFT $\mathrm{Sym}^N(T^4)$ has for all practical purposes been derived~\cite{Eberhardt:2018ouy,Eberhardt:2019ywk}. In the hope to find generalisations, the understanding of this correspondence is however still undergoing rapid development. Not too long ago, a certain free field realisation of the affine algebra $\mathfrak{psu}(1,1|2)_{k=1}$ (which enters the world-sheet description of the string theory in the hybrid formalism~\cite{Berkovits:1999im}) was for instance introduced to streamline aspects of the derivation~\cite{Dei:2020zui}. In particular, the free field realisation was used to efficiently demonstrate the fact that the path integral of the world-sheet model localises onto field configurations that represent covering maps $\Gamma : \Sigma \rightarrow S^2$, where $\Sigma$ is the world-sheet and  $S^2$ is the conformal boundary of $\mathrm{AdS}_3$ in Euclidean signature.

A crucial aspect of the free field realisation introduced in~\cite{Dei:2020zui} was an emergent twistorial interpretation of the world-sheet model, which was then further elaborated in~\cite{McStay:2024dtk}. The algebra $\mathfrak{psu}(1,1|2)_{k=1}$ was expressed as the BRST reduction of two pairs of symplectic bosons, denoted in this article by $X_i$, $Y^i$, $i\in \{1,2\}$, and two pairs of free fermions ($\chi_i$, $\xi^i$, $i\in \{1,2\}$) by the current
\begin{equation}\label{eq:BRST}
    \mathcal{J} = X_i Y^i - \chi_i \xi^i~.
\end{equation}
With $\Gamma (z)$ the covering map, the following schematic relation for physical correlators was found
\begin{equation}\label{eq:classical-loc}
    \langle X_1 (z) - \Gamma (z)  X_2 (z) \rangle_{\mathrm{phys}} = 0~,
\end{equation}
which is reminiscent of the ``incidence relations" $x_1 -\gamma x_2=0$ satisfied by twistor ($x_1,x_2$) and  standard ($\gamma$) variabes for the conformal boundary.~\footnote{Alternatively, in this degenerate case of twistor space, one may simply identify $x_1$ and $x_2$ with homogenous coordinates on the boundary sphere, which is isomorphic to its twistor space as a complex manifold. The two perspectives are often rather harmlessly conflated in the literature, \emph{c.f.}~\cite{Dei:2020zui,Dei:2023ivl}, and we will do so here too. The difference becomes relevant in higher dimensions.} In turn, the bosonic free field variables $X_i$, $Y^i$ were then naturally interpreted (by analogy with Berkovits' twistor string~\cite{Berkovits:2004hg}) as chiral avatars of projective coordinates $(x_i,y^i)$ on $(\mathbb{P}^1)^* \times \mathbb{P}^1$, the ambitwistor space of the boundary $S^2 \cong \mathbb{P}^1$.~\footnote{For further discussions on these ambitwsitor spaces as well as their relations to mini-twistor spaces for $\mathrm{AdS}_3$, see \emph{e.g.}~\cite{Adamo:2016rtr,McStay:2024dtk, Bu:2023cef}.} Leveraging this intuition, a free field realisation of $\mathfrak{psu}(2,2|4)_1$ was then introduced by simply doubling the variables in~\eqref{eq:classical-loc} (that is, taking $i\in \{1,\cdots ,4\}$), which were interpreted as functions on a quadric in $(\mathbb{P}^3)^* \times \mathbb{P}^3$ defining the ambitiwstor space of compactified/complexified four dimensional Minkowski space. This has opened the way to a proposal for the string theory dual to free large-$N$ $\mathcal{N}=4$ SYM~\cite{Gaberdiel:2021jrv}.

A closely related family of Vertex Operator Algebras (VOAs), based on the Lie algebras $\mathfrak{psl}(n|n)$ for any integer $n>1$, has meanwhile appeared in the context of boundary vertex algebras of $A$-twisted 3d $\mathcal{N}=4$ theories. These are the simple quotients of affine $\mathfrak{psl}(n|n)$ VOAs at level $1$, which in accordance with standard mathematical literature we will denote as $V_1(\mathfrak{psl}(n|n))$, and which for $n=2$ and $n=4$ are essentially the compexifications of the affine algebras considered above. In particular, the author of this article first conjectured together with Chris Beem~\cite{Beem:2023dub}, and then proved with Aiden Suter~\cite{Ferrari:2024dst}, that the \emph{associated variety} of $V_1(\mathfrak{psl}(n|n))$ for integer $n>1$ is  $\bar{\mathbb{O}}_{\mathrm{min}}(\mathfrak{sl}_n^*)$, the minimal nilpotent orbit closure in in $\mathfrak{sl}_n^*$. The motivation behind these studies was the formulation of an analogue of the \emph{Higgs branch conjecture}~\cite{Beem:2017ooy} --a pivotal element of the 4d SCFT/2d VOA correspondence~\cite{Beem:2013sza} that identifies the \emph{Higgs branch of the moduli space of vacua} of a 4d $\mathcal{N}=2$ SCFTs with the \emph{associated variety} of the respective VOA-- in the context of the 3d TQFT/2d boundary VOA correspondence. In fact, the 3d $\mathcal{N}=4$ theories SQED[$n$] were conjectured in~\cite{Costello:2018fnz} (essentially based on the free field realisation later introduced in~\cite{Gaberdiel:2021jrv}, although from a different perspective) to support some quotient of affine $\mathfrak{psl}(n|n)$ at level $1$ at their boundary. In~\cite{Beem:2023dub}, it was conjectured that the quotient is in fact the simple quotient, and arguments for this statement where given in the $n=2$ case. The Higgs branches of $\mathrm{SQED}[n]$ are the minimal nilpotent orbit closures, which can be defined in terms of non-chiral versions of the BRST reduction~\eqref{eq:BRST}. In~\cite{Ferrari:2024dst} it was indeed proven that these are the associated varieties of $V_1(\mathfrak{psl}(n|n))$.

This statement implies, by definition and in a rigorous sense~\cite{arakawa2012remark}, that the VOAs $V_1(\mathfrak{psl}(n|n))$ are \emph{chiral quantisations} of the algebra of functions on these minimal nilpotent orbit closures $\bar{\mathbb{O}}_\mathrm{min} (\mathfrak{sl}(n,\mathbb{C})^*)$. Notice that this can also na\"ively be interpreted as a ``localisation" statement in the holographic sense, in that without quotienting by null vectors one would expect the associated varieties to be isomorphic to $\mathfrak{sl}(n,\mathbb{C})^*\times \mathfrak{sl}(n,\mathbb{C})^*$ (the space which even elements in $\mathfrak{psl}(n|n)$ are functions on). In the simple quotients, the associated varieties become $\bar{\mathbb{O}}_\mathrm{min} (\mathfrak{sl}(n,\mathbb{C})^*) \cong \{pt.\}\times \bar{\mathbb{O}}_\mathrm{min} (\mathfrak{sl}(n,\mathbb{C})^*) \subset \mathfrak{sl}(n,\mathbb{C})^*\times \mathfrak{sl}(n,\mathbb{C})^*$ and are therefore much smaller. Now it is well known that there are resolutions of singularities (the so-called Springer resolutions)
\begin{equation}\label{eq:springer}
    T^*\mathbb{P}^{n-1} \rightarrow \omin~,
\end{equation}
and so $V_1(\mathfrak{psl}(n|n))$ can equally well be understood as chiral quantisations of these varieties.~\footnote{Strictly speaking, when working with the resolution, one would need to chirally quantise the structure sheaf of these smooth varieties and take global sections to recover the VOAs. This would be a super-extension of the work~\cite{kuwabara2021vertex}, which is in progress. For similar developments, see~\cite{arakawa2023hilbert,Coman:2023xcq}.}
Thus, in view of the intuitive reasons provided for the localisation of the path integral of the world-sheet model mentioned above, this immediately raises the following interesting question:
\begin{itemize}
    \item[\bf{Q1}:]{Are $V_1(\mathfrak{psl}(2|2))$, $V_1(\mathfrak{psl}(4|4))$ chiral quantisations of varieties related to the conformal boundaries of complexified $\mathrm{AdS}_3$, $\mathrm{AdS}_5$?}
\end{itemize}
And if yes, more ambitiously:
\begin{itemize}
    \item[\bf{Q2}:]{Is the observed localisation of the path integral of minimal tension string theory on $\mathrm{AdS}_3\times S^3\times T^4$, in some meaningful sense, a consequence of this fact?}
\end{itemize}

The first purpose of this article is to give some evidence that the answers to~\textbf{Q1}~and~\textbf{Q2} are affirmative. In fact, good evidence for \textbf{Q1} in the $\mathrm{AdS}_3$ context can already be found by carefully comparing statements in the aforementioned literature. Free field realisations for $V_1(\mathfrak{psl}(2|2))$, $V_1(\mathfrak{psl}(4|4))$ that implement the reduction by~\eqref{eq:BRST} automatically (thus ``solving" the BRST reduction problem at hand) were systematically written down in~\cite{Beem:2023dub}. These were \emph{a posteriori} interpreted (in the spirit of~\cite{Beem:2019tfp}) as localisation procedures on openly embedded subsets of the associated variety
\begin{equation}\label{eq:asso-loc}
   (T^*\mathbb{C}^*)^{m-1} \times (T^*\mathbb{C})^{n-m-1}  \hookrightarrow T^*\mathbb{P}^{n-1}~.
\end{equation}
Here, following~\cite{gorbounov1999gerbes}, we identify free fields (half-lattice vertex algebras/$\beta\gamma$-systems) with chiralisations of functions on~$T^*\mathbb{C}^*$/$T^*\mathbb{C}^*$, their associated varieties. Now for $n=2$ one of the free field realisations of~\cite{Beem:2023dub} were utilised in~\cite{Dei:2023ivl} to further streamline the proof of the path integral localisation on $\mathrm{AdS}_3$, and after a careful analysis of the world-sheet model the fields were interpreted as functions on 
\begin{equation}\label{eq:hol-loc}
 \mathbb{C} \hookrightarrow \mathbb{P}^1    
\end{equation}
where $\mathbb{P}^1$ is the (twistor space of) the Euclidean conformal boundary. In this article we observe (Section~\ref{sec:free-field-2}) that the ``localisations"~\eqref{eq:asso-loc}~and~\eqref{eq:hol-loc} are indeed compatible, thus corrobotating an affirmative answer to~\textbf{Q1}: the core of the associated variety can be identified with the twistor space of the boundary sphere, and the associated variety is its cotangent bundle. The associated variety, however, differs from the standard interpretation of the free field variables in the holographic context in other ways; for instance, the cotangent bundle $T^*\mathbb{P}^1$ includes the zero section (which is clearly absent in $(\mathbb{P}^1)^*\times \mathbb{P}^1$). We leave the inspection of this subtle difference to future work.

Although answering the likely overambitious, and less precise, \textbf{Q2} is admittedly more subtle, the following can at least be observed. One crucial fact of the localisation argument presented in~\cite{Dei:2023ivl} is the existence of an operator $D$ (related to the ``secret representations" of~\cite{Eberhardt:2019ywk}) that allegedly ensures that certain free fields are actually functions on $\mathbb{P}^1$  (and not simply $\mathbb{C}\subset \mathbb{P}^1)$. In this article we explain (Section~\ref{subsec:D2}) how the operator $D$ can be constructed systematically as a screening operator, and that its role is simply to ensure that its kernel defines the simple quotient $V_1(\mathfrak{psl}(2|2))$ inside the free field algebra.

The second purpose of this article is then, given this evidence, to assume the relation between resolutions of the associated variety of $V_1(\mathfrak{psl}(4|4))$ with the cotangent bundle of the twistor space $\mathbb{P}^3$ of the conformal boundary of $\mathrm{AdS}_5 \times S^5$, and to explicitly produce a free field realisation for $V_1(\mathfrak{psl}(4|4))$ that
\begin{enumerate}
    \item Unlike the one provded in~\cite{Gaberdiel:2021jrv}, does not require any quotient by any current
    \item The free fields can manifestly be interpreted as chiralisations of functions on $\mathbb{P}^3$.
\end{enumerate}
A free field realisation satisfying 1. was already produced in~\cite{Beem:2013sza} (in fact for all $n>0$), but since the motivation was different it was based on subsets of the \emph{singular} minimal nilpotent orbit closure appearing in~\eqref{eq:springer}, and not its resolution (and so subsets not supported on the exceptional divisor $\mathbb{P}^3$). Thus, we explicitly write down an alternative free field realisation that satisfies this requirement, based on other choices to be made in the framework of~\cite{Beem:2023dub}. This has the advantage that it makes an economical use of screening operators, the $n=4$ cousins of $D$, needing one instead of two, which we write down systematically. We hope that this free field realisation can find applications in the study of the proposed world-sheet dual to free $\mathcal{N}=4$ SYM in the large-$N$ limit.

This article is structured as follows. In Section~\ref{sec:ass-var} we recall basic facts on associated varieties of VOAs, and summarise some of the results we will need in this domain. In Section~\ref{sec:wakimoto} we recall some classic results on Wakimoto free field realisations that will be needed to justify the identication of the associated varieties with cotangent bundles of twistor spaces. In Section~\ref{sec:free-field-2} we discuss the free field realisations of $V_1(\mathfrak{psl}(2|2))$, give evidence in favour of this identification, and briefly explain the origin of the screening operators $D$. In Section~\ref{sec:free-field4} we discuss similar free field realisations for $V_1(\mathfrak{psl}(4|4))$, enjoying the two features highlighted above.

\paragraph{Acknowledgements.} The author would like to thank C.~Beem, A.~Dei, B.~Knighton, V.~Schomerus and A.~Suter for discussions.

\section{Preliminary remarks on associated varieties}
\label{sec:ass-var}

We start by making some preliminary remarks on the notion of~\emph{associated variety}. We emphasise the by now well-known idea that a VOA may be the thought of as a \emph{chiral quantisation} of some algebra of functions on a space. For universal affine VOAs based on a simple Lie algebra $\mathfrak{g}$, these varieties are isomorphic to $\mathfrak{g}^*$ whereas for quotients these are generically subvarieties of $\mathfrak{g}^*$. Thus, in some sense, depending on the level the simple quotients of affine VOAs can be ``localised" on subvarieties of $\mathfrak{g} ^*$.

\subsection{Definition}

We first recall the definition of associated variety of a VOA $V$~\cite{arakawa2012remark}. We assume the VOA to be strongly finitely generated, with strong generators $\lbrace a^i,\dots,a^r\rbrace$ (this is the case for the affine VOAs considered here). Define~\footnote{Here and below we use standard mathematics notations, where for a field $a$, $a_{(n)}$ corresponds to the the mode of a field expanded as $a(z) = \sum_{n\in \mathbb{Z}} a_{(n)}z^{-n-1}$.}
\begin{equation}
    C_2(V)=\mathrm{span}_\mathbb{C} \lbrace a^i_{(-n)}v\mid 1\leq i\leq r,\; n\geq 2,\; v\in V\rbrace~.
\end{equation}
In other words, $C_2(V)$ is the span over $\mathbb{C}$ of those concatenations of strong generators acting on the vacuum where at least one of the fields is differentiated. The Zhu $C_2$ algebra of $V$ is defined to be
\begin{equation}\label{eq:C2}
    R_V= V/C_2(V)~,
\end{equation}
and has a commutative product given by
\begin{equation}
    \overline{a}\cdot \overline{b}=\overline{a_{(-1)}b}~,
\end{equation}
for $ a,b\in V$, where the bar denotes the equivalence class. The associated variety of $V$ is defined as
\begin{equation}
    X_V =\mathrm{specm}( R_V)~.
\end{equation}
$R_V$ has furthermore the structure of a Poisson algebra, with bracket is given by
\begin{equation}
    \lbrace\cdot,\cdot\rbrace:R_V\times R_V \to R_V~,~
    (\overline{a},\overline{b}) \mapsto \overline{a_{(0)}b}~.
\end{equation}
It follows that the associated variety has the structure of a Poisson variety. A VOA $V$ is by definition a \emph{chiral quantisation} of the algebra of functions on $X_V$. In favourable cases, such as for VOAs that are classically free, there is a unique quantisation procedure that recovers a VOA $V$ from the arc space of $X_V$ (or rather its scheme-theoretic, unreduced counterpart). Although some special cases of classically free VOAs are known~\cite{creutzig2024classical}, which affine VOAs are chirally free remains a wildly open problem.

\subsection{Affine case}

Assume that $\mathfrak{g}$ is a simple Lie algebra (we will only be interested in $\mathfrak{g}=\mathfrak{sl}(n,\mathbb{C})$, $n>1$, but it is easy to state some results more generally). Then it is well-known that for the \emph{universal} affine VOA $V^k(\mathfrak{g})$, we have~(see \cite{arakawa2021arc} and references therein)
\begin{equation}
    R_V = \mathrm{Sym}(\mathfrak{g})
\end{equation}
and so 
\begin{equation}
    X_{V^k (\mathfrak{g})} = \mathfrak{g}^*~,
\end{equation}
Thus, in one can view the universal affine VOA $V^k(\mathfrak{g})$ as a \emph{chiral quantisation} of the algebra of functions on $\mathfrak{g}^*$. Notice incidentally that the Poisson variety $\mathfrak{g}^*$ has infinitely many symplectic leaves (the co-adjoint orbits). This means that conjecturally, it cannot be identified with the Higgs branch of supersymmetric quantum field theories.

If we have some quotient  $V^k(\mathfrak{g})/U$ of $V^k(\mathfrak{g})$ by some submodule $U$, then we will have in general
\begin{equation}
    R_V \cong  \mathrm{Sym}(\mathfrak{g}) / I
\end{equation}
where $I$ is some ideal in $ \mathrm{Sym}(\mathfrak{g})$. Thus, the associated variety will be a subvariety of $\mathfrak{g}^*$, and more precisely a $\mathfrak{g}$-invariant subvariety of $\mathfrak{g}^*$. If this subvariety is contained in the nilcone, it has the chance to be the Higgs branch of some supersymmetric theory.

\subsection{$V^{k}(\mathfrak{sl}(n,\mathbb{C}))$ and some quotients}

Consider now the special case of the universal affine VOA $V^k(\mathfrak{sl}(n,\mathbb{C}))$. By the above, we have that
\begin{equation}
    X_{V^k(\mathfrak{sl}(n,\mathbb{C}))} \cong \mathfrak{sl}(n,\mathbb{C})^* ~.
\end{equation}
For any positive integer $k>0$, let $V_{k>0}(\mathfrak{sl}(n,\mathbb{C}))$ be the simple quotient of the affine VOA. Then one can show that all elements in  $R_V$ are nilpotent, essentially because
\begin{equation}\label{eq:nil}
    e_{\theta , (-1)}^{k+1} \ket{0} \sim 0~,
\end{equation}
where $e_\theta$ is the highest root of $\mathfrak{sl}(n,\mathbb{C})$ and $\ket{0}$ is the vacuum, is singular and thus set to zero in the quotient. The $\mathfrak{sl}(n,\mathbb{C})$ action then implies that $h_{(-1)}$ and $f_{(-1)}$ are also nilpotent. It follows that the associated variety is a point
\begin{equation}
    X_{V_k(\mathfrak{sl}(n,\mathbb{C}))} \cong\{ pt\}~.
\end{equation}

\subsection{$V^{k}(\mathfrak{psl}(n|n))$ and some quotients}
\label{subsec:Vkasso}
We can now consider the case $V^{k}(\mathfrak{psl}(n|n))$. Since odd elements are nilpotent and the even subalgebra is
\begin{equation}
    V^{-k} (\mathfrak{sl}(n,\mathbb{C})) \times V^k (\mathfrak{sl}(n,\mathbb{C}))~, 
\end{equation}
one can immediately conclude
\begin{equation}
    X_{V^{k}(\mathfrak{psl}(n|n))} \cong  X_{V^k (\mathfrak{sl}(n,\mathbb{C}))} \times X_{V^k(\mathfrak{sl}(n,\mathbb{C}))} \cong  \mathfrak{sl}(n,\mathbb{C})^* \times \mathfrak{sl}(n,\mathbb{C})^*~.
\end{equation}
For $k=1$, moreover, it was proved in~\cite{Ferrari:2024dst} that
\begin{equation}
    X_{V_{1}(\mathfrak{psl}(n|n))} \cong \bar{\mathbb{O}}_{\mathrm{min}}(\mathfrak{sl}(n,\mathbb{C})^*)\times \{pt\}~,
\end{equation}
where $\bar{\mathbb{O}}_{\mathrm{min}}(\mathfrak{sl}(n,\mathbb{C})^*)$ is the minimal nilpotent orbit closure in $\mathfrak{sl}(n,\mathbb{C})^*$. This is (essentially) because the maximal submodules of $V^{\pm 1} (\mathfrak{sl}(n,\mathbb{C}))$ can be shown to be contained in the maximal submodule of $V^1(\mathfrak{psl}(n|n))$, and so by a generalisation of~\eqref{eq:nil}
\begin{equation}
    X_{V_{1}(\mathfrak{psl}(n|n))} \subset \mathfrak{sl}(n,\mathbb{C})^* \times \{pt\}~.
\end{equation}
Furthermore, all $2\times 2$ minors in $\mathfrak{sl}(n,\mathbb{C})$ can be shown to be nilpotent in $R_{V_{1}(\mathfrak{psl}(n|n))}$.\footnote{In the special case $n=2$, it is sufficient to note that there is a conformal embedding of the even subalgebra into the full VOA; equating the Sugawara constructions for the even subalgebra with the Sugawara stress tensor of the full VOA shows that there is some null relation that sets the Casimir of $V_{-1}(\mathfrak{sl}(2,\mathbb{C}))$ to be proportional to an odd element. See~\cite{Beem:2023dub} for more details.} 

At higher but finite $k$, we still have
\begin{equation}
     X_{V_{k}(\mathfrak{psl}(n|n))} \subset \mathfrak{sl}(n,\mathbb{C})^* \times \{pt\}~.
\end{equation}
In the semi-classical limit $k\rightarrow \infty$ the associated variety tends to the full variety $\mathfrak{sl}(n,\mathbb{C})^*\times \mathfrak{sl}(n,\mathbb{C})^*$.

\subsection{Affine $\mathfrak{psl}(n|n)$ and the 3d SQED[$n$] Higgs branch theorem}
\label{subsec:VOA-Higgs-th}

We now briefly recall the 3d SQED[$n$] Higgs branch/VOA theorem~\cite{Ferrari:2024dst}, as we will need it below. In general, the Higgs branch conjecture~\cite{Beem:2017ooy} in the context of the 4d $\mathcal{N}=2$ SCFTs/VOA correspondence states that the associated variety of the a VOA emerging from a 4d $\mathcal{N}=2$ SCFT is isomorphic as a Poisson variety to the Higgs branch of its moduli space of vacua. A similar conjecture was formulated  in~\cite{Beem:2023dub} in the context of the A-twisted 3d $\mathcal{N}=4$ theories/2d VOA correspondence, and substantiated in several ways in the case of abelian gauge theories. We now explain how this works in one of the most elementary cases, 3d $\mathcal{N}=4$ SQED[$n$].

3d $\mathcal{N}=4$ SQED[$n$] is a $U(1)$ gauge theory with $n$ hypermultiplets in the fundamental representation. The bottom components of these hypermultiplets can be taken to be functions $(x_i , y^i ) \in \mathbb{C}[T^*\mathbb{C}]$, $i\in \{1,\cdots n\}$, and the gauge group acts with unit weight $+ 1$ on $x_i$ (and $-1$ on $y^i$). 

\paragraph{Higgs branch.} The Higgs branch is defined as the hyperk\"ahler reduction of the hyperk\"ahler space $T^*\mathbb{C}^n$ by the action of the gauge group. Fixing a complex structure on $T^*\mathbb{C}^n$, the Higgs branch can be viewed as a holomorphic symplectic space 
\begin{equation}
M_H(n) \cong T^*\mathbb{C}^n /\! /\! / U(1) \cong \mu_{\mathbb{C}}^{-1}(0) /\! /\mathbb{C}^*~,
\end{equation}
where $\mathbb{C}^*$ is a complexification of the gauge group and
\begin{equation}
    \mu_{\mathbb{C}} = y^ix_i
\end{equation}
is the complex moment map of the gauge action on $T^*\mathbb{C}^n$. By defintion, it is the spectrum of
\begin{equation}
    R_H(n) = \mathbb{C}[\mu_{\mathbb{C}}^{-1}(0)]^\mathbb{C^*}~,
\end{equation}
We can package these functions into a rank-1 $n\times n$ matrix
\begin{equation}\label{eq:M-pres}
M = \begin{pmatrix}
    x_1y^1 & \cdots & x_1 y^n \\
    \vdots   &\cdots & \vdots             \\
    x_n y^1 &\cdots  & x_n y^n
\end{pmatrix}~.
\end{equation}
On the zero locus of the moment map $\mu_{\mathbb{C}} = y^ix_i = 0$, $M$ squares to zero. Thus, the spectrum of the space of such matrices is isomorphic to the minimal nilpotent orbit closure $\bar{\mathbb{O}}_{\mathrm{min}}(\mathfrak{sl}(n,\mathbb{C})^*)$.

This variety can be resolved by means of a character of $\mathbb{C} ^*$, say $ \zeta \in \mathbb{Z}^+$
\begin{equation}\label{eq:res}
M_{H,\zeta}(n) \cong \mu_{\mathbb{C}}^{-1}(0) /\! /_\zeta \mathbb{C}^*~\cong T^*\mathbb{P}^{n-1}~.
\end{equation}
One useful way to understand this resolution is in terms of yet another moment map construction, as follows. Consider the action of $\mathrm{SL}(n,\mathbb{C})$ on $T^*\mathrm{SL}(n,\mathbb{C})$, endowed with the standard symplectic form coming from the cotangent bundle structure, defined by left multiplication. It is hamiltonian, and has moment map
\begin{equation}
 \mu_{\mathbb{C}}^{S} :  T^*\mathrm{SL}(n,\mathbb{C}) \rightarrow \mathfrak{sl}(n,\mathbb{C})^*~.
\end{equation}
Now we have $\mathbb{P}^n\cong \mathrm{SL}(n,\mathbb{C})/P$ where $P$ is a parabolic subgroup. The $\mathrm{SL}(n,\mathbb{C})$ action descends to this quotient and has moment map
\begin{equation}\label{eq:Paction}
 \widetilde{\mu_{\mathbb{C}}^{S}} :  T^*\mathbb{P}^{n-1} \rightarrow \mathfrak{sl}(n,\mathbb{C})^*~,
\end{equation}
whose image is the minimal nilpotent orbit closure
\begin{equation}
    \mathrm{Im}\left(\widetilde{\mu_{\mathbb{C}}^S}\right) \cong \bar{\mathbb{O}}_{\mathrm{min}}(\mathfrak{sl}(n,\mathbb{C})^*)~.
\end{equation}

\paragraph{Boundary VOA.} The boundary VOA of SQED[$n$] was identified in~\cite{Costello:2018fnz} with a chiral analogue of this symplectic reduction, but with extra odd copies of the fields. Let $X_i$, $Y^i$ be symplectc bosons pairs
\begin{equation}
    X_i(z) Y^j (w) \sim \frac{\delta_i^j}{z-w}~,
\end{equation}
representing boundary values of the 3d fields $x_i$, $y^i$. Let $i,j\in \{1,\cdots , n\}$, and introduce boundary free fermion pairs
\begin{equation}
    \chi_i (z) \xi^j  (w) = \frac{\delta_i^j}{z-w}~,
\end{equation}
with again $i,j \in \{1, \cdots n\}$. The VOA generated by these pairs contains a level-0, rank-1 Heisenberg current
\begin{equation}
    \mathcal{J} = Y^iX_i + \xi^j\chi_j~,
\end{equation}
and the boundary VOA of A-twisted SQED[$n$] was identified with the relative BRST reduction of this free VOA by the current $\mathcal{J}$. Notice that $\mathcal{J}$ is a chiral analogue of the complex moment map $\mu_\mathbb{C}$, where the boundary fermions are introduced to cancel gauge anomalies (in more mathematical terms, the BRST current becomes level zero).

It was already suggested in~\cite{Costello:2018fnz} that the VOA is some quotient of $V^1(\mathfrak{psl(n|n)})$, whose representatives we take to be of the schematic form (for $i,j\in \{1,\cdots , n\}$)
\begin{equation}\label{eq:schamatic}
    \begin{pmatrix}
        V^{-1} (\mathfrak{sl}(n,\mathbb{C})) &     S^+ \\
                            S^-        &    V^{1} (\mathfrak{sl}(n,\mathbb{C}))
    \end{pmatrix}       
    =
    \begin{pmatrix}
        \{X_iY^j\}_i^{\ j} & \{ X_i \xi^{\ j} \}_i^{\ j} \\
        \{\chi_i Y^j\}_i^{ \ j} &\{ \chi_i \xi^{\ j} \}_i^{\ j}
    \end{pmatrix}~.
\end{equation}
For $n=2$ this is essentially (up to a complexification) the free field realisation suggested in~\cite{Dei:2023ivl} for affine $\mathfrak{psu}(1,1|2)$ at level $1$. It was argued in~\cite{Beem:2023dub} for $n=2$ at a physical level of rigour, and proven in~\cite{Ferrari:2024dst} for $n>2$, that the result is indeed the simple quotient $V_1(\mathfrak{psl}(n|n))$. Thus, combining this with the results in Section~\ref{sec:ass-var}, in~\cite{Ferrari:2024dst} a proof was given of the fact that the associated variety is isomorphic to the Higgs branch as a Poisson variety. This is the what we call the 3d SQED[$n$] Higgs branch/VOA theorem~\cite{Ferrari:2024dst}.

\section{Wakimoto constructions}
\label{sec:wakimoto}

As reviewed above, the associated variety of $V_1(\mathfrak{psl}(2|2))$ is isomorphic (as a Poisson variety) to $\bar{\mathbb{O}}_{\mathrm{min}}(\mathfrak{sl}(2,\mathbb{C})^*)\times \{pt\}$. Given the proposed connection to minimal tension string theory, we would like to interpret this as a ``holographic" statement. That is, although one would expect the isometries of complexified $\mathrm{AdS}_3\times S^3$ to be valued in functions on $\mathfrak{sl}(2,\mathbb{C})^*\times \mathfrak{sl}(2,\mathbb{C})^*$, null relations at $k=1$ imply that the associated variety of the vertex algebra only receives contributions from a subvariety related to the boundary of the $\mathrm{AdS}_3$ factor.~\footnote{As reviewed in the previous section, there are nilpotent contributions that come from the $S^3$ part. It is natural to view these as Kaluza-Klein modes.}

To pin down the precise relation to the boundary, one handy approach is to determine how free fields of $V_1(\mathfrak{psl}(2|2))$ relate to space-time variables. One standard way to do this in holography is via the Wakimoto construction~\cite{deBoer:1998gyt,Giveon:1998ns,Kutasov:1999xu}, which we quickly review here. For more details on minimal tension see, for instance~\cite{Dei:2023ivl,McStay:2024dtk} and~\cite{McStay:2023thk} for the complexified setting.

\subsection{Classical considerations}

Consider complexified $\mathrm{AdS}_3$, which we view as the hyperboloid
\begin{equation}
    \mathrm{AdS}_3 = \lbrace ( z_0 ,z_1, z_2, z_3 )  \in \mathbb{C}^4~|~ \sum_{i=0}^3{z^2_i} = 1 \rbrace ~,
\end{equation}
with metric induced from the ambient metric $\sum_{i=0}^{3}dz_i^2$. We can introduce some standard complex coordinates $\Phi$, $\boldsymbol{\gamma}$ and $\tilde{\boldsymbol{\gamma}}$\footnote{In Euclidean signature $\tilde{\boldsymbol{\gamma}}$ is the complex conjugate of $\boldsymbol{\gamma}$ and $\Phi$ is real. Then $g\in H^+_3 \cong \mathrm{SL}(2,\mathbb{C})/\mathrm{SU}(2)$. In Lorentzian signature, $\Phi$, $\boldsymbol{\gamma}$ and $\tilde{\boldsymbol{\gamma}}$ are all independent real variables.}
\begin{equation}
    e^{\Phi} = z_0 - iz_3~,~
    \boldsymbol{\gamma} = \frac{-z_1 + iz_2}{z_0-iz_3}~,~\tilde{\boldsymbol{\gamma}} = \frac{z_1+iz_2}{z_0-iz_3}~.
\end{equation}
Conventions are chosen so that we can package these variables into a matrix $g\in \mathrm{SL}(2,\mathbb{C})$
\begin{equation}\label{eq:quot-rep}
    g = e^{\Phi} 
\begin{pmatrix}
e^{-2\Phi}+\boldsymbol{\gamma} \tilde{\boldsymbol{\gamma}}  &  \boldsymbol{\gamma} \\
\tilde{\boldsymbol{\gamma}} & 1
\end{pmatrix}~,
\end{equation}
and identify $\mathrm{AdS}_3 $ with $\mathrm{SL}(2,\mathbb{C})$. The ``conformal boundary" $g_\infty$ of complexified of $\mathrm{AdS}_3$ can be obtained by taking the limit $\mathfrak{Re}(\Phi) \rightarrow \infty$
\begin{equation}\label{eq:adsbound}
    g_\infty = \lim_{\mathfrak{Re}(\Phi) \rightarrow \infty} g =
    e^{\Phi}
    \begin{pmatrix}
        \boldsymbol{\gamma} \tilde{\boldsymbol{\gamma}}  &  \boldsymbol{\gamma} \\
    \tilde{\boldsymbol{\gamma}} & 1
    \end{pmatrix}~.
\end{equation}
Thus, the boundary is naturally parametrised by the holomorphic (in the large-radius limit) coordinate $\boldsymbol{\gamma}$.

Consider now the holomorphic conserved currents constructed out of matrix representatives of the form~\eqref{eq:quot-rep}. We can compute
\begin{equation}
    J = -( \partial g) g^{-1} = 
    \begin{pmatrix}
       -e^{2\Phi} \boldsymbol{\gamma} \partial \tilde{\boldsymbol{\gamma}} + \partial \Phi & e^{2\Phi} \boldsymbol{\gamma}^2 \partial \tilde{\boldsymbol{\gamma}} -\partial \boldsymbol{\gamma} - 2 \boldsymbol{\gamma} \partial \Phi \\
       -e^{2\Phi} \partial \tilde{\boldsymbol{\gamma}} & e^{2\Phi} \boldsymbol{\gamma} \partial \boldsymbol{\gamma} - \partial \Phi
    \end{pmatrix}~,
\end{equation}
which we rewrite as 
\begin{equation}\label{eq:WZW}
   J = e^{2\Phi}
    \begin{pmatrix}
       -\boldsymbol{\gamma}\partial{\tilde{\boldsymbol{\gamma}}} + \partial \Phi e^{-2\Phi} \quad  & \boldsymbol{\gamma}^2\partial{\tilde{\boldsymbol{\gamma}}} -\partial \boldsymbol{\gamma} e^{-2\Phi}  - 2 \boldsymbol{\gamma} \partial \Phi e^{-2\Phi}  \\
        - \partial{\tilde{\boldsymbol{\gamma}}} &  \boldsymbol{\gamma} \partial{\tilde{\boldsymbol{\gamma}}} - \partial \Phi e^{-2\Phi} 
    \end{pmatrix}~.
\end{equation}
This should be understood as a parametrisation of functions on $\mathfrak{sl}(2,\mathbb{C})^*$. We introduce for future convenience the conjugate variable $\boldsymbol{\beta} = -e^{2\Phi } \partial \tilde{\boldsymbol{\gamma}} $, so that we can rewrite~\eqref{eq:WZW} as follows
\begin{equation}\label{eq:JWaki}
   J = 
    \begin{pmatrix}
       \boldsymbol{\gamma}\boldsymbol{\beta}+ \partial \Phi  \quad  & -\boldsymbol{\gamma}^2\boldsymbol{\beta}-\partial \boldsymbol{\gamma}   - 2 \boldsymbol{\gamma} \partial \Phi   \\
        \boldsymbol{\beta}&   -\boldsymbol{\gamma} \boldsymbol{\beta}- \partial \Phi
    \end{pmatrix}~.
\end{equation}

\subsection{The Wakimoto free field realisation}
\label{subsec:Wakimoto}

We recall that the matrix~\eqref{eq:JWaki} is the foundation of the Wakimoto free field realisation, which reads
\begin{align}\label{eq:Waki}
    e &= \beta \\
    f &=  -( \beta \gamma ) \gamma +   \partial \Phi \gamma  \\
    h &=  2\beta \gamma - \partial \Phi~.\label{eq:Waki3}
\end{align}
where we have promoted the variables $\boldsymbol{\beta}$ and $\boldsymbol{\gamma}$ to a symplectic boson pair $(\beta,\gamma)$
\begin{equation}
 \beta (z) \gamma (z) \sim \frac{-1}{z-w}~,
\end{equation}
and $\Phi$ is a free boson
\begin{equation}
    \Phi (z) \Phi (w)  \sim \frac{1}{(z-w)^2}~.
\end{equation}
The missing term in $f$ with respect to~\eqref{eq:JWaki} is a quantum effect, which depends on the level. It is clear from this perspective that $\gamma$ chiralises the boundary sphere. In the next section, we will review another free field realisation that will be compared with the Wakimoto one.

\section{Free field realisations of $V_1(\mathfrak{psl}(2|2))$}
\label{sec:free-field-2}

We now summarise (in a simplified fashion) the free field realisation for $V_1(\mathfrak{psl}(2|2))$ obtained in~\cite{Beem:2023dub}, and then utilised in~\cite{Dei:2023ivl} in the context of the $\mathrm{AdS}_3/\mathrm{CFT}_2$ correspondence. We first explain what is the underlying geometric interpretation in terms of the associated variety, and then study the free field realisation proper. The final identification of the free fields will corroborate an affirmative answer to~\textbf{Q1} in the introduction. In particular, we demonstrate that the core $\mathbb{P}^1$ of the resolved associated variety is parametrised by homogenous coordinates that satisfy the incidence relations.

We will furthermore emphasise how, by exploiting the techniques of~\cite{Beem:2023dub}, the auxiliary operator $D$ of~\cite{Dei:2023ivl} (a re-conceptualisation of the ``secret representation" of~\cite{Eberhardt:2018ouy}) can be systematically constructed and interpreted. In some sense, it merely ensures that one is dealing with a quotient (argued in~\cite{Beem:2023dub} to be the simple quotient) of $V^1(\mathfrak{psl}(2|2))$ inside the free field algebra. The systematic construction is an aspect that was missing in~\cite{Dei:2023ivl}. Since this operator plays a crucial role in the path integral localisation argument of~\cite{Dei:2023ivl}, the construction provides minimal evidence for a positive answer to \textbf{Q2}. It will also be important in view of the higher-rank generalisation we will consider the next section.

\subsection{Finite dimensional considerations, localisations and twistor spaces}
\label{subsec:P1}

As mentioned in the introduction, we expect free field realisations to emerge from a chiral analogue of localisations on associated varieties. Set as before $n=2$. Then recall that the non-zero gauge-invariant elements in $R_H(2)$ are
\begin{equation}
    e = x_1y^2~, ~f=x_2y^1~,~h =1/2( x_1y^1 - x_2 y^2)~.
\end{equation}
We remarked that these can be packaged into an $\mathfrak{sl}(2,\mathbb{C})$ matrix $M$ that is nilpotent.
\begin{equation}
    M=\begin{pmatrix}
        \frac{1}{2}(x_1y^1 - x_2y^2 ) & x_1 y^2 \\
        x_2 y^1 &  -\frac{1}{2}(x_1y^1 - x_2y^2 ) 
    \end{pmatrix}~,~\mathrm{det}(M)=M^2=0~. 
\end{equation}
If we resolve the minimal nilpotent orbit to $T^*\mathbb{P}^1$ as in~\eqref{eq:res}, then $x_1$ and $x_2$ can be identified with homogenous coordinates on $\mathbb{P}^1$.

We can obtain interesting subsets of the associated variety by making either $x_i$ or $y^i$ invertible, for $i \in \{1,2\}$~\cite{Beem:2023dub}. If we set $x_1, x_2 \neq 0 $ (the $\epsilon = (++)$ choice in~\cite{Beem:2023dub}), we get a subset 
\begin{equation}\label{eq:fin-non-deb}
T^*\mathbb{C}^* \hookrightarrow T^*\mathbb{P}^1~,
\end{equation}
where $\mathbb{C}^* \hookrightarrow \mathbb{P}^1$. The homomorphism is straightforward~\cite{Beem:2023dub},  functions on $\mathbb{C}^*$ are the $x_1^{-1} x_{2}$ and its inverse. We could then extend this subset to include the origin of $\mathbb{C}^*$ 
\begin{equation}\label{eq:fin-deb}
    T^*\mathbb{C} \hookrightarrow T^*\mathbb{P}^1~,
\end{equation}
where now $\mathbb{C} \hookrightarrow \mathbb{P}^1$.~\footnote{Notice that as explained in~\cite{Beem:2023dub} we could consider subsets where $x_1 \neq 0$, $y^2 \neq 0$ instead (the $\epsilon = (+-)$ choice of~\cite{Beem:2023dub}). Then we would get a subset $T^*\mathbb{C}^* \hookrightarrow \bar{\mathbb{O}}_{\mathrm{min}}(\mathfrak{sl}(2,\mathbb{C})^*)$ whose functions are $e,h$, and which is not supported on the exceptional divisor $\mathbb{P}^1$. We will not consider this case here.} Our free fields will be based on these open subsets, and we will see in the next section that a chiralisation of the coordinate $x_1^{-1}x_2$ will be identified with the field $\gamma$ in the Wakimoto construction. This is the statement at the origin of the ``incidence relation"~\eqref{eq:classical-loc} that twistor variables for the boundary are supposed to satisfy, and which in~\cite{Dei:2020zui} was taken to be the hallmark of the path integral localisation.

It however instructive to already ask, at this classical and rather superficial level, whether resolution variables $x_1$, $x_2$ can in some way be identified with with ``twistor variables" for the boundary of $\mathrm{AdS}_3$. To do so in an admittedly imprecise way, consider the parametrisation of $\mathfrak{sl}(2,\mathbb{C})^*$ provided by the Wakimoto variables~\eqref{eq:JWaki}. There is an obvious parametrisation of $\bar{\mathbb{O}}_{\mathrm{min}}(\mathfrak{sl}(2,\mathbb{C})^*)\subset \mathfrak{sl}(2,\mathbb{C})^* $ that can be obtained from this by neglecting derivatives of the fields $\boldsymbol{\gamma}$ and $\boldsymbol{\Phi}$
\begin{equation}
   M= \begin{pmatrix}
        \boldsymbol{\gamma} \boldsymbol{\beta} & - \boldsymbol{\gamma}^2 \boldsymbol{\beta}\\
     \boldsymbol{\beta}&  -\boldsymbol{\gamma}  \boldsymbol{\beta}
    \end{pmatrix}~,~\mathrm{det}(M)=M^2=0~.
\end{equation}
When considering the associated variety one needs to neglect derivatives of generators (as recalled around~\eqref{eq:C2}), so this step does not appear unreasonable. It would however require further justification, involving for instance nilpotency of the operator $\partial \Phi$ at $k=1$. For now we will assume it is justified, and that the above matrix is a reasonable image of the generators expressed in the free field variables in $R_V$. Then we want relate these variables to the variables $x_1$, $x_2$ and $y^1$, $y^2$, remembering that $x_1$, $x_2$ can be viewed as homogenous coordinates on the exceptional divisor $\mathbb{P}^1$ of the Springer resolution $T^* \mathbb{P}^1$. To do so, notice that we must identify
\begin{equation}\label{eq:id1}
    \boldsymbol{\beta} = x_2 y^1
\end{equation}
as well as
\begin{equation}\label{eq:id2}
    \boldsymbol{\gamma} \boldsymbol{\beta} = x_1y^1~.
\end{equation}
Thus, if we divide~\eqref{eq:id2}~by~\eqref{eq:id1}, we obtain 
\begin{equation}\label{eq:incidence-cl}
x_2 -  \boldsymbol{\gamma}  x_1 = 0`~,
\end{equation}
This is the classical analogue of the incidence relation.

Notice that the ``derivation" presented here is quite different from those proposed in the literature~\cite{McStay:2024dtk}. The latter start from natural properties of twistor variables, take a large radius limit, and then use these twistor variables to parametrise WZW currents precisely in the above way $M=\{x_iy^j\}_i^j$. From our point of view this is quite unsatisfactory, because it does not explain why currents are taken to be rank-1 when parametrised in terms of these twistorial variables. It is nonetheless natural to ask whether the full set of variables $x_i$, $y^i$ can be identified with the ambitwistor space of the boundary $\mathbb{P}^1$, which are closely related~\cite{Adamo:2016rtr,Bu:2023cef}. That is, one can consider the ambitwistor space of $\mathbb{P}^1$
\begin{equation}\label{eq:ambi}
    (\mathbb{P}^1)^*  \times \mathbb{P}^1
\end{equation}
parametrised by variables $\tilde y_i$, $\tilde x_i$ (say) respectively but independently, oppositely projectivised. From our point of view, it is however better to think of our variables as functions on $T^*\mathbb{P}^1$, including the zero section. Moreover, we do really want to impose the complex moment map condition $\mu_{\mathbb{C}}=0$, which enforced in addition to~\eqref{eq:ambi} defines what in the literature is known as ``Euclidean" ambitwistor space~\cite{McStay:2024dtk}. We will leave an exploration of this subtle difference to future work.

\subsection{Bosonisation and screening}
\label{subsec:22bos}

To produce a free field realisation of this quotient, we consider a chiral analogue of the inversion formulae studied in the previous section. This is the set of bosonisations
\begin{equation}
    (X_i , Y^i) \mapsto (e^{\rho_i - \sigma_i}, \partial {\rho_i} e^{-\rho_i +\sigma_i})
\end{equation}
where $\langle \rho_i,\rho_j \rangle=\delta_{i,j}$, $\langle \sigma_i,\sigma_j \rangle=-\delta_{i,j}$, as well as 
\begin{equation}
    (\xi^{i} , \chi_i) \mapsto (e^{\nu_i}, e^{-\nu_i})
\end{equation}
with $\langle \nu_i, \nu_j \rangle = \delta_{i,j}$. Notice that $X_1$, $X_2$ have been made ``invertible". The image of the symplectic boson pairs~\cite{allen2022bosonic} into half-lattice extensions for Heisenberg currents $\partial \rho_i$, $\partial \sigma_i$ is given by the kernel of the screening operator
\begin{equation}
    \mathfrak{S}_i = \mathrm{Res}_{z=0} \mathfrak{s}_i~,
\end{equation}
where
\begin{equation}\label{eq:2scr}
    \mathfrak{s}_i = e^{\rho_i}~.
\end{equation}
We will make use of a fairly common abuse of language below and call~\eqref{eq:2scr} the screening operator.

\subsection{Change of variables and free field realisation}

To simplify the BRST problem, we may consider the following change of variables~\cite{Beem:2023dub}
\begin{equation}
\begin{alignedat}{1}
\eta &= \sigma_1 + \sigma_2 +\nu_1 + \nu_2 \\
\tilde{\eta} &= \sigma_1 + \sigma_2 -\rho_1 - \rho_2 \\
\phi &= -\sigma_1 + \sigma_2 \\
\delta &= \rho_1 -\rho_2 \\
\omega_1 &= \frac{1}{2}(\sigma_1 + \sigma_2) - \frac{1}{2}(\rho_1 +\rho_2) + \nu_1 \\
\omega_2 &=  \frac{1}{2}(\sigma_1 + \sigma_2) - \frac{1}{2}(\rho_1 +\rho_2) + \nu_2~.
\end{alignedat}
\end{equation}
The current $\mathcal{J}$ determining the BRST operator becomes
\begin{equation}
    \mathcal{J} = \partial \eta
\end{equation}
and its only non-trivial OPEs are with $\tilde{\eta}$. Thus, the solution to the BRST problem can be written in terms of the free fields $\delta$, $\phi$, $\omega_1$ and $\omega_2$: BRST-closed operators will be independent of $\tilde{\eta}$ by construction, and terms involving $\partial \eta$ will be exact.

One key observation is that
\begin{equation}\begin{alignedat}{1}
\partial \rho_ 1 &= \frac{1}{2}(\partial \eta + \partial \delta + \psi_1\tilde{\psi}^1+\psi_2 \tilde{\psi}^2) \\
\partial \rho_ 2 &= \frac{1}{2}(\partial \eta - \partial \delta + \psi_1\tilde{\psi}^1+\psi_2 \tilde{\psi}^2)~.
\end{alignedat}\end{equation} 
Since $\partial \eta$ is BRST-exact, unique representatives of these operators in BRST cohomology are
\begin{equation}\begin{alignedat}{1}
\bar{J}_{\rho_ 1} &= \frac{1}{2}(\partial \delta -\partial \omega_1 - \partial \omega_2) \\
\bar{J}_{\rho_ 2} &= \frac{1}{2}( - \partial \delta - \partial \omega_1 - \partial \omega_2)~.
\end{alignedat}\end{equation}
Similarly. the screening operators~\eqref{eq:2scr} acting on free-field space become
\begin{equation}\begin{alignedat}{1}
\mathfrak{s}_1 &= e^{ \frac{1}{2} ( \delta -\omega_1 - \omega_2 )} \\
\mathfrak{s}_2 &= e^{ \frac{1}{2} ( -\delta -\omega_1 - \omega_2)}~.
\end{alignedat}\end{equation}

By definition, they must commute with the $V_1(\mathfrak{psl}(2|2))$ generators (as their kernel defines the image of the symplectic boson embedding, and the generators are built out of symplectic bosons).

Let us write down the $V_1(\mathfrak{psl}(2|2))$ generators in the free-field variables. We begin with
\begin{equation}\begin{alignedat}{3}
& e &&= X_1Y^2  &&=  \bar{J}_{\rho_2} e^{\phi+\delta}    \\
& f &&= X_2 Y^1 && =    \bar{J}_{\rho_1} e^{-\phi-\delta}   \\
& h &&= X_1Y^2 - X_2Y^1 &&= - \partial \phi~.
\end{alignedat}\end{equation}
Then we have
\begin{equation}\begin{alignedat}{3}
& (S^+)_1^{\ 1}  &&= X_1\xi^1  &&=  e^{\frac{1}{2}(\phi+\delta )} \tilde{\psi}^1    \\
& (S^+)_1^{\ 2} &&= X_1\xi^2  &&=   e^{\frac{1}{2}(\phi+\delta) }  \tilde{\psi}^2  \\
& (S^+)_2^{\ 1} &&= X_2\xi^1  &&=   e^{\frac{1}{2}(\phi+\delta)}   \tilde{\psi}^1 \\
& (S^+)_2^{\ 2} &&= X_2\xi^2  &&=   e^{\frac{1}{2}(\phi+\delta)}   \tilde{\psi}^2 \\
\end{alignedat}\end{equation}
as well as
\begin{equation}\begin{alignedat}{3}
& (S^-)_1^1 &&= \chi_1Y^1  &&=  \bar{J}_{\rho_1} e^{-\frac{1}{2}(\phi+\delta)} \psi_1   \\
& (S^-)_1^2 &&= \chi_1Y^2  &&=  \bar{J}_{\rho_2} e^{-\frac{1}{2}(\phi+\delta)}  \psi_2  \\
& (S^-)_2^1 &&= \chi_1Y^1  &&=  \bar{J}_{\rho_1} e^{-\frac{1}{2}(\phi+\delta)}   \psi_1  \\
& (S^-)_2^2 &&= \chi_2Y^2  &&=    \bar{J}_{\rho_2}  e^{-\frac{1}{2}(\phi+\delta)} \psi_2    ~.
\end{alignedat}\end{equation}
It is straightforward to write down the fermion bilienars in the new variables.

\subsection{De-bosonisation and the ``secret" operator $D$}
\label{subsec:D2}

Above we have obtained two screening operators, but we can get rid of one of them by ``de-bosonisation". This procedure is the chiral analogue of going from~\eqref{eq:fin-non-deb}~to~\eqref{eq:fin-deb}. In fact, we may observe that the pair
\begin{equation}
   (\gamma , \beta ) :=  (e^{-\phi-\delta} , \bar{J}_{\rho_{2}}e^{\phi+\delta} ) 
\end{equation}
satisfies symplectic bosons OPEs. Moreover, the screening operator $\mathfrak{s}_2$ is precisely the screening operator that determines the image of $(\gamma , \beta)$ into free-field space. Thus, we may want to de-bosonise these fields, work directly with these $\gamma$ and $\beta$ and forget about the screening operator $\mathfrak{s}_2$. Geometrically, then, $\gamma$ is as anticipated the chiralisation of a coodinate on
\begin{equation}
    \mathbb{C} \hookrightarrow \mathbb{P}^1~,
\end{equation}
where $\mathbb{P}^1$ is the exceptional divisor of the Springer resolution. As we shall see, the repeated use of the $\gamma$ symbol is an informative a notational abuse; it will in fact be identified with the field $\boldsymbol{\gamma}$ parametrising the boundary.

To de-bosonise, however, we actually need to trade the bosonised fermions $e^{\pm \omega_1 }$ and $e^{\pm \omega_2}$ for an alternative pair of free fermions. This is because we must be able to write $h=-\partial \phi$ as well as $\bar{J}_{\rho_1}$ in terms of $\gamma$, $\beta$ as well as the pairs of fermions. It is clearly not possible to do so with the chosen $e^{ \pm \omega_i}$. However, if we set
\begin{equation}\begin{alignedat}{2}
    \Psi_i &= e^{\frac{1}{2} (\delta+\phi)} \psi_i \\
    \tilde{\Psi}_i &= e^{-\frac{1}{2} (\delta+\phi)} \tilde{\psi}^i~,
\end{alignedat}\end{equation}
then we can express
\begin{equation}\begin{alignedat}{2}
        \partial \phi &= -2\beta \gamma + \Psi_i \tilde{\Psi}^i                      \\
        \bar{J}_{\rho_1} &= -\beta \gamma + \Psi_i\tilde{\Psi}^i~.
\end{alignedat}\end{equation}
Thus, we can systematically rewrite all generators of $V_1(\mathfrak{psl}(2|2))$ in terms of these fields. For instance,
\begin{equation}\begin{alignedat}{3}
& e &&=  \bar{J}_{\rho_2} e^{\phi+\delta}   &&= \beta  \\
& f && =    \bar{J}_{\rho_1} e^{-\phi-\delta}  &&=  ( -\beta \gamma + \Psi_i\tilde{\Psi}^i) \gamma   \\
& h&&= -\partial \phi  &&= 2\beta \gamma - \Psi_i \tilde{\Psi}^i~.
\end{alignedat}\end{equation}
Expressing the other generators in terms of the new free-fields is a simple exercise that we leave to the reader.

The result manifestly agrees with the free field realisation used in~\cite{Dei:2023ivl} to prove the localisation of the path integral of the $\mathrm{AdS}_3\times S^2 \times T^4$ world-sheet model. Moreover, by bosonising the fermions $\Psi^i$ we then obtain the Wakimoto free-field realisation~\eqref{eq:Waki}-\eqref{eq:Waki3}~\cite{Dei:2023ivl}. We can as a consequence make a couple of remarks. First, in the Wakimoto construction the field $\gamma$ was interpreted as a chiral version of the coordinate $\boldsymbol{\gamma}$ parametrising the (twistor space of the) boundary of $\mathrm{AdS}_3$. That this is actually the case can also be shown by an explicit computation of its transormation under boundary translations. Thus, we may indeed identify the exceptional divisor $\mathbb{P}^1$ with the twistor space of the boundary, corroborating an affirmative answer to~\textbf{Q1}.
\begin{itemize}
    \item The field $\gamma$ chiralising a functions on a Zariski dense subset of the core of the associated variety is identified with a chiralisation of the coordinate $\boldsymbol{\gamma}$ parametrising a Zariski dense subset of the (twistor space of the) boundary of $\mathrm{AdS}_3$. The associated variety is therefore identified with the cotangent bundle of (the twistor space of) the boundary.
\end{itemize}
Second, we may turn our attention to the screening operator $\mathfrak{s}_1$ and see whether it is related to the operator $D$ of~\cite{Dei:2023ivl}. Recall that $\mathfrak{s}_1$ reads
\begin{equation}
    \mathfrak{s}_1 = e^{\frac{1}{2}(\delta -\omega_1 - \omega_2)}~.
\end{equation}
Notice that its OPE with $\gamma$ produces a pole,
\begin{equation}
    \gamma (z) \mathfrak{s}_1 (w) \sim \mathcal{O}\left( \frac{1}{z-w} \right)~,
\end{equation}
as follows from
\begin{equation}
   \gamma =  e^{-\delta - \phi} ~,~ \langle \delta , \delta \rangle = 2~.
\end{equation}
Also, recall that by construction it commutes with all of the $\mathfrak{psl}(2|2)$ generators. These were the defining properties of the operator $D$ introduced in~\cite{Dei:2023ivl}, see Section~4.1,~eq.~4.2 to complete the proof of the localisation of the path integral. Thus, it can be concluded that the operator $D$ is a screening operator whose kernel determines some quotient of affine $\mathfrak{psl}(2|2)$ at level $1$ into free field space, corroborating a (minimal) affirmative answer to~\textbf{Q2}. Notice incidentally that this operator \emph{cannot} explicitly be written down in terms of the new free fields, if not in a formal way (in~\cite{Dei:2023ivl} it was denoted as a delta function).

\section{Free field realisations of $V_1(\mathfrak{psl}(4|4))$} 
\label{sec:free-field4}

In this section we focus on the proposal of~\cite{Gaberdiel:2021jrv} for the world-sheet model of minimal tension string theory in $\mathrm{AdS}_5$, which is based on $\mathfrak{psu}(2,2|4))_{k=1}$. As in the $\mathrm{AdS}_3$ case, we consider the simple quotient of the complexified affine VOA $V_1(\mathfrak{psl}(4|4))$ and construct a suitable free field realisation.

As mentioned in the introduction, the free field realisation we derive is modelled on functions on $T^*\mathbb{P}^3 \rightarrow \bar{\mathbb{O}}_\mathrm{min}(\mathfrak{sl}_4^*)$, where $\mathbb{P}^3$ following~\cite{Gaberdiel:2021jrv} should be interpreted as the twistor space of the boundary. Moreover, much like the free field realisation for $V_1(\mathfrak{psl}(2|2))$, the final free field realisation does not require any BRST quotient. However, due to the obscure space-time origin of the world-sheet model, a direct interpretation of the minimal nilpotent orbit in terms of a boundary limit of space-time currents is currently lacking, and we leave this to future work. 

\subsection{Finite dimensional analogy and functions on $T^*\mathbb{P}^3$}

To construct free field realisations, the same steps as in the $V_1(\mathfrak{psl}(2|2))$ case can essentially be carried out for $V_1(\mathfrak{psl}(4|4))$. In fact, similar free field realisations of $V_1(\mathfrak{psl}(n|n))$ for all integer $n>1$ were considered in~\cite{Beem:2023dub}. Unfortunately, these were relying on a bosonisation modelled on the geometry of the minimal nilpotent orbit closure $\bar{\mathbb{O}}_{\mathrm{min}}(\mathfrak{sl}_4^*)$, and the free field variables were not supported on $\mathbb{P}^3\subset T^* \mathbb{P}^3$ of its Springer resolution~\eqref{eq:res}
\begin{equation}
    T^*\mathbb{P}^3 \rightarrow \bar{\mathbb{O}}_{\mathrm{min}}(\mathfrak{sl}_4^*) ~.
\end{equation}
We therefore proceed to write down a free field realisation whose variables are indeed supported on this $\mathbb{P}^3$, and which unlike the one utilised in~\cite{Gaberdiel:2021jrv} does not involve any quotient.

The desired free field realisation will be based, in the language of~\cite{Beem:2023dub}, on the bosonisation $\epsilon = (++++)$, which means that it is modelled on the following finite dimensional analogy. In the above symplectic resolution the variables $x_i$, $i\in \{1,\cdots ,4\}$ introduced in Section~\ref{subsec:VOA-Higgs-th} descend to homogenous coordinates on $\mathbb{P}^3$, and in particular not all of them can be simultaneously set to zero. If we restrict ourselves to a subset where all of them are simultaneously non-zero, then we obtain a subset
\begin{equation}
    T^* (\mathbb{C}^*)^3 \hookrightarrow T^* \mathbb{P}^3~.
\end{equation}
where again $(\mathbb{C}^*)^3$ is an openly embedded subset of $\mathbb{P}^3$. We can also consider function that extend to some extra points to land on the subset
\begin{equation}
    T^* \mathbb{C}^3 \hookrightarrow T^* \mathbb{P}^3
\end{equation}
instead. Below we will perform similar operations at the level of vertex algebras.

\subsection{Bosonisation and screening}
\label{subsec:44bos}

Recall from Section~\ref{subsec:VOA-Higgs-th} that for SQED[$4$] we have to  this case we take four symplectic boson pairs
\begin{equation}
    X_i(z) Y^j (w) \sim \frac{\delta_i^j}{z-w}~,
\end{equation}
with $i,j\in \{1,4\}$, as well as two free fermion pairs
\begin{equation}
    \chi_i (z) \xi^j  (w) = \frac{\delta_i^j}{z-w}~,
\end{equation}
with again $i,j \in \{1,4\}$. It was proved in~\cite{Ferrari:2024dst} that the relative BRST reduction of the VOA generated by these pairs by the level-0, rank-1 Heisenberg current
\begin{equation}
    \mathcal{J} = Y^iX_i + \xi^j\chi_j
\end{equation}
(where Einstein summation convention is understood) is the simple quotient $V_1(\mathfrak{psl}(4|4))$, whose associated variety is a minimal nilpotent orbit closure in $\mathfrak{sl}(4,\mathbb{C})^*$. This reduction coincides with the one proposed in~\cite{Gaberdiel:2021jrv} in the context of minimal tension string theory in $\mathrm{AdS}_5$. 

To produce a free field realisation of this simple quotient that ``solves"" the BRST reduction, we repeat the same steps as in Section~\ref{subsec:44bos}. We consider
\begin{equation}\begin{alignedat}{2}
    (X_i , Y^i) &\mapsto (e^{\rho_i - \sigma_i}, \partial {\rho_i} e^{-\rho_i +\sigma_i}) \\
    (\xi^{i} , \chi_i) &\mapsto (e^{\nu_i}, e^{-\nu_i})
\end{alignedat}\end{equation}
We now have \emph{four} screening operators
\begin{equation}
    \mathfrak{s}_i = e^{\rho_i}~.
\end{equation}
We will be able to get rid off one of them via de-bosonisation techniques, but one will still act as an analogue of the operator $D$.

\subsection{Change of basis}

We now consider a generalised change of variables based on~\cite{Beem:2023dub}. First, we introduce a boson modelled on the BRST current and an additional boson that pairs non-trivially with it
\begin{equation}\begin{alignedat}{2}
\eta = \sum_{i=1}^4\sigma_i +\nu_i ~ , \quad && \tilde{\eta} = \sum_{i=1}^4\sigma_i +\rho_i~ .
\end{alignedat}\end{equation}
Second, we introduce six bosons that have vanishing OPEs with the preceding ones
\begin{equation}\begin{alignedat}{2}
\phi_1 &= -\sigma_1 + \sigma_2~,  \quad && \delta_1 = \rho_1 - \rho_2 \\
\phi_2 &= -\sigma_1 + \sigma_3~ ,  \quad && \delta_2 = \rho_1 - \rho_3 \\
\phi_3 &= -\sigma_1 + \sigma_4~ ,  \quad && \delta_3 = \rho_1 - \rho_4 ~ .
\end{alignedat}\end{equation}
Third, we introduce the following ``fermionic" directions
\begin{equation}\begin{alignedat}{2}
\omega_1 &= \frac{1}{4} \left( \sum_{j=1}^4 \sigma_j - \rho_j \right)+ \nu_1~, \quad  && \omega_2 = \frac{1}{4}\left( \sum_{j=1}^4 \sigma_j - \rho_j \right) + \nu_2 \\
\omega_3 &= \frac{1}{4} \left( \sum_{j=1}^4 \sigma_j - \rho_j \right) + \nu_3~ , \quad &&\omega_4 = \frac{1}{4}\left( \sum_{j=1}^4 \sigma_j - \rho_j \right)+ \nu_4~.
\end{alignedat}\end{equation}

Notice that now 
\begin{equation}
    \langle \delta_i , \delta_j \rangle = \begin{pmatrix}
        2 & 1 & 1 \\
        1 & 2 & 1 \\
        1 & 1 & 2
    \end{pmatrix}~,
\end{equation}
and similarly (but with a negative sign) for the $\phi_i$.

As before, the key observation for the purposes of this article is that the BRST problem is essentially concentrated in the $\eta$, $\tilde{\eta}$ variables. We further observe that
\begin{equation}\begin{alignedat}{1}
 \rho_1 &= \frac{1}{4}\left( \eta + \sum_{i=1}^3  \delta_i  - \sum_{j=1}^4 \omega_j   \right) \\
 \rho_2 &= -\delta_1 + \frac{1}{4}\left( \eta + \sum_{i=1}^3  \delta_i  - \sum_{j=1}^4 \omega_j   \right) \\
 \rho_3 &= -\delta_2 + \frac{1}{4}\left( \eta + \sum_{i=1}^3  \delta_i  - \sum_{j=1}^4 \omega_j   \right) \\
 \rho_4 &= -\delta_3 + \frac{1}{4}\left( \eta + \sum_{i=1}^3  \delta_i  - \sum_{j=1}^4 \omega_j   \right) ~.
\end{alignedat}\end{equation}

This allows us to write down the represnetatives of the currents $\partial \rho_i$ in the BRST cohomology
\begin{equation}\begin{alignedat}{1}
 \bar{J}_{\rho_1} &= \frac{1}{4}\partial\left(\sum_{i=1}^3  \delta_i  - \sum_{j=1}^4 \omega_j   \right) \\
 \bar{J}_{\rho_2} &= -\partial \delta_1 + \frac{1}{4}\partial\left( \sum_{i=1}^3  \delta_i  - \sum_{j=1}^4 \omega_j   \right) \\
\bar{J}_{\rho_3} &= -\partial \delta_2 + \frac{1}{4}\partial \left(  \sum_{i=1}^3  \delta_i  - \sum_{j=1}^4 \omega_j   \right) \\
 \bar{J}_{\rho_4} &= -\partial\delta_3 + \frac{1}{4}\partial\left( \sum_{i=1}^3  \delta_i  - \sum_{j=1}^4 \omega_j   \right) ~,
\end{alignedat}\end{equation}
as well as the screening operators $\mathfrak{s}_i$
\begin{equation}\begin{alignedat}{1}
\mathfrak{s}_1 &= e^{\frac{1}{4}\left( \sum_{i=1}^3  \delta_i  - \sum_{j=1}^4 \omega_j   \right)} \\
\mathfrak{s}_2 &= e^{-\delta_1 + \frac{1}{4}\left( \sum_{i=1}^3  \delta_i  - \sum_{j=1}^4 \omega_j   \right) } \\
\mathfrak{s}_3 &= e^{- \delta_2 + \frac{1}{4} \left( \sum_{i=1}^3  \delta_i  - \sum_{j=1}^4 \omega_j   \right)} \\
 \mathfrak{s}_4 &= e^{-\delta_3 + \frac{1}{4}\left( \sum_{i=1}^3  \delta_i  - \sum_{j=1}^4 \omega_j   \right) }~.
\end{alignedat}\end{equation}

It is once again straightforward to write down the $V_1(\mathfrak{psl}(4|4))$ generators in the new variables. Let us consider the $V_{-1}(\mathfrak{sl}(4))$ subalgebra, which arises from the symplectic boson pairs in~\eqref{eq:schamatic}
\begin{equation}\label{eq:4-boson}
\{ X_iY^j \}_i^j \mapsto 
    \begin{pmatrix}
        \star & \bar{J}_{\rho_2} e^{-(\phi_1+\delta_1)}  & \bar{J}_{\rho_3} e^{-(\phi_2+\delta_2)}   & \bar{J}_{\rho_4} e^{-(\phi_3+\delta_3)} \\
        \bar{J}_{\rho_1} e^{(\phi_1+\delta_1)}  &\star  & \bar{J}_{\rho_3} e^{(\phi_1+\delta_1)-(\phi_2+\delta_2)}   & \bar{J}_{\rho_4} e^{(\phi_1+\delta_1)-(\phi_3+\delta_3)} \\
        \bar{J}_{\rho_1} e^{(\phi_3+\delta_3)} & \bar{J}_{\rho_2} e^{(\phi_1+\delta_1)-(\phi_2+\delta_2)}  & \star   & \bar{J}_{\rho_4} e^{(\phi_2+\phi_2) - (\phi_3+\delta_3)} \\
        \bar{J}_{\rho_1} e^{(\phi_3+\delta_3)} & \bar{J}_{\rho_2} e^{(\phi_1+\delta_1)-(\phi_3+\delta_3)}  & \bar{J}_{\rho_3} e^{(\phi_2+\delta_2)-(\phi_3+\delta_3)}   & \star ~,
    \end{pmatrix}
\end{equation}
whereas the Cartan elements are linear combinations of $\partial {\phi_1} $, $ \partial {\phi_2}$, $\partial {\phi_3}$. The $V_1(\mathfrak{sl}(4))$ subagebra is generated by fermionic bilinears, whereas odd elements can be written down as in the $n=2$ case.

\subsection{Debosonisation and the operator $D$}

We can get rid off the three screening operators $\mathfrak{s}_2$, $\mathfrak{s}_3$, $\mathfrak{s}_4$  by debosonisation. The screening operator
\begin{equation}\label{eq:D4}
    \mathfrak{s}_1 = e^{\frac{1}{4}\left(  \sum_{i=1}^3  \delta_i  - \sum_{j=1}^4 \omega_j   \right)}
\end{equation}
will remain as a higher dimensional analogue of the operator/``secret representation" $D$.

To debosonise, notice that the three pairs
\begin{equation}\begin{alignedat}{1}
(\beta_1,\gamma^1)&= \left( \bar{J}_{\rho_2} e^{-(\phi_1+\delta_1)}  , e^{(\phi_1+\delta_1)} \right) \\
(\beta_2,\gamma^2)&= \left( \bar{J}_{\rho_3} e^{-(\phi_2+\delta_2)}  , e^{(\phi_2+\delta_2)} \right) \\
(\beta_3,\gamma^3)&= \left( \bar{J}_{\rho_4} e^{-(\phi_3+\delta_3)}  , e^{(\phi_3+\delta_3)} \right)~.
\end{alignedat}\end{equation}
satisfy the symplectic boson OPEs
\begin{equation}
  \gamma_i (z) \beta^j (w) = \frac{\delta_i^j}{z-w}~.
\end{equation}
Moreover, the image of the $\gamma^i$ and $\beta_j$ into the free field VOA is precisely the kernel of the screening operators we want to get rid of.

In order to build a free field realisation of $V_1(\mathfrak{psl}(4|4))$ in terms of the debosonised variables, however, we need to be able to express (as before) $\bar{J}_{\rho_1}$ as well as $\partial \phi_i$ (which generate the Cartan) in terms of these variables as well as four free fermions. It is again possible if we consider
\begin{equation}\begin{alignedat}{1}
\Psi_i &= e^{\frac{1}{4}\sum_{i=1}^3 (\delta_j + \phi _j)} \psi_i \\
\tilde{\Psi}^i &= e^{-\frac{1}{4}\sum_{i=1}^3 (\delta_j + \phi _j)} \tilde{\psi}^i ~.
\end{alignedat}\end{equation}
Then we have
\begin{equation}
    \bar{J}_{\rho_1} = - \sum_{i=1}^3 \beta_i \gamma^i + \sum_{j=1}^4 \Psi_j \tilde{\Psi}^j~,
\end{equation}
as well 
\begin{equation}\begin{alignedat}{1}
    - 4  \partial \phi_i +\sum_{j=1}^3 \partial \phi_j = -4 \beta_i \gamma^i + \sum_{i=1}^{4} \Psi_i \tilde{\Psi}^i~.
\end{alignedat}\end{equation}
Solving for $\partial \phi_i$ (a simple linear problem) we can obtain the Cartan elements generating $V_{-1}(\mathfrak{sl}(4))$.

The other elements of this subalgebra (those entering the matrix~\eqref{eq:4-boson}) can then be expressed as 
\begin{equation}
\label{eq:fee-field-5}
    \begin{pmatrix}
        \star & \beta_1 & \beta_2 & \beta_3 \\
        \bar{J}_{\rho_1}\gamma^1  &\star  & \gamma^1\beta_2   & \gamma^1\beta_3 \\
        \bar{J}_{\rho_1} \gamma^2 & \gamma^2 \beta_1  & \star   & \gamma^2 \beta_3\\
        \bar{J}_{\rho_1} \gamma^3 & \gamma^3\beta_1 & \gamma^3 \beta_2   & \star ~.
    \end{pmatrix}~.
\end{equation}
The subalgebra generated by fermionic bilinars is easy to write down
\begin{equation}
        \begin{pmatrix}
        \star & \Psi_1 \tilde{\Psi}^2 & \Psi_1 \tilde{\Psi}^3 & \Psi_1 \tilde{\Psi}^4 \\
        \Psi_2 \tilde{\Psi}^1 &\star  & \Psi_2 \tilde{\Psi}^3 & \Psi_2 \tilde{\Psi}^4 \\
         \Psi_3 \tilde{\Psi}^1  & \Psi_3 \tilde{\Psi}^2  & \star   & \Psi_3 \tilde{\Psi}^4\\
        \Psi_4 \tilde{\Psi}^1 &  \Psi_4 \tilde{\Psi}^2  &  \Psi_4 \tilde{\Psi}^3   & \star ~.
    \end{pmatrix}~,
\end{equation}
with obvious Cartan generators.

The odd generators read as follows:
\begin{equation}\begin{alignedat}{2}
    X_1 \xi^j &= \tilde{\Psi}^j   &&  \quad Y^1 \chi_j = \bar{J}_{\rho_1} \Psi_j \\
    X_2 \xi^j &= \gamma_1 \tilde{\Psi}^j  && \quad    Y^2 \chi_j = \beta_1 \Psi_j  \\
    X_3 \xi^j &= \gamma_2 \tilde{\Psi}^j &&  \quad  Y^3 \chi_j = \beta_2  \Psi_j  \\
    X_4 \xi^j &= \gamma_3 \tilde{\Psi}^j &&  \quad  Y^4 \chi_j = \beta_3 \Psi_j~. 
\end{alignedat}\end{equation}
We hope that this free field realisation, together with the screening operator~\eqref{eq:D4}, will help to shed light onto the string theory dual of free large-$N$ $\mathcal{N}=4$ SYM. We leave this to future work.

\appendix

\bibliographystyle{JHEP}
\bibliography{associated_varieties}

\end{document}